\documentclass{aa}
\usepackage{psfig}

\def\bron{1RXS J171824.2--402934}
\def\ecs{erg~cm$^{-2}$s$^{-1}$}

\begin{document}
\thesaurus{08.09.2 \bron; 08.14.1; 13.25.1}

\title{Discovery of \bron\ as an X-ray burster}

\author{R.G.~Kaptein\inst{1}
 \and  J.J.M.~in~'t~Zand\inst{1}
 \and  E.~Kuulkers\inst{1,2}
 \and  F.~Verbunt\inst{2}
 \and  J.~Heise\inst{1}
 \and  R.~Cornelisse\inst{1,2}
}

\offprints{R.G.Kaptein (ronaldk@sron.nl)}

\institute{     Space Research Organization Netherlands, Sorbonnelaan 2,
                NL - 3584 CA Utrecht, the Netherlands 
	\and
                Astronomical Institute, Utrecht University, P.O. Box 80000,
                NL - 3508 TA Utrecht, the Netherlands
	}

\date{Received, accepted }

\maketitle 

\begin{abstract}

We report the first-time detection of a type-I X-ray burst from a
position that is consistent with that of \bron. The detection was made
on Sep 23, 1996, with the Wide Field Cameras on board the BeppoSAX
satellite. The burst had a peak intensity of 1.3 Crab units in 2 to 28
keV and is relatively long (at least 3.5~min). Analysis of the burst
gives clear evidence for photospheric radius expansion. Assuming the
peak luminosity to be close to the Eddington limit, standard burst
parameters and taking into account gravitational redshift effects, the
distance to the source is approximately 6.5 kpc. No other bursts from
this source have been observed during the rest of the WFC
observations. The detection of a type-I burst implies that \bron\ in
all likelihood is a low-mass X-ray binary where the compact object is
a neutron star.

\end{abstract}

\section{Introduction}
\label{intro}
\bron\ was first detected during the ROSAT All-Sky Survey (Voges et
 al. 1999) on September 2, 1990, when it had a count rate of 0.157 c
 s$^{-1}$ (0.1-2.4 keV) in the Position Sensitive Proportional
 Counter. The spectral hardness of the source was high, suggesting an
 X-ray binary (Motch et al. 1998). Two pointed observations were done
 with ROSAT in March and September 1994 with net exposure times of 790
 and 2479 seconds respectively using the High-Resolution Imager.  The
 count rates were 0.048 and 0.019 c s$^{-1}$ respectively. The
 position was determined to be $\alpha_{2000.0}=17^{\rm h}18^{\rm
 m}24.13^{\rm s}, \delta_{2000.0}=-40\degr 29\arcmin 30\farcs4$ with a
 90\% confidence error radius of 15\farcs7 (Motch et al. 1998).  Motch
 et al. searched for the optical counterpart, but were unable to
 identify one.

We here present the analysis of a short flare that was observed from
the same sky position with the Wide Field Cameras (WFCs) on the
BeppoSAX satellite in September 1996 and explain it within the
framework of thermonuclear flashes on neutron stars.

\section{Observations}
\label{observations}
The WFCs (Jager et al. 1997) are two identical coded aperture cameras
on board the BeppoSAX satellite (Boella et al. 1997), which was
launched on April 30, 1996. Each camera consists of a coded mask and a
position-sensitive proportional counter. The field of view is
$40^{\degr}\times40^{\degr}$ (full width to zero response), the
angular resolution 5\arcmin\ and the source location accuracy is
generally better than 1\arcmin. The energy range is 2 to 28 keV with
an energy resolution of 18$\%$ at 6 keV. The on-axis detection
threshold is of the order of a few mCrab in $10^{5}$~s, and varies as
a function of the total flux contained in the field of view. The WFCs
are particularly suitable for the study of fast transient X-ray
phenomena at unexpected sky positions, thanks to the large coverage of
the sky in combination with good angular resolution.

In 1996 the Galactic center was targeted by the WFCs for 22 days
during August 21 to October 29. The source reported here is 13$\fdg$4
from the Galactic center, well within the field of view. The net
exposure time of the source during this period is 550 ks.

\section{Detection and position}
\label{detection}

One of the WFCs detected the X-ray flare on Sep 23.313, 1996 (UT, or
MJD 50349.313) at a position of $\alpha_{2000.0} = 17^{\rm h}18^{\rm
m}23.6^{\rm s}, \delta_{2000.0} = -40\degr29\arcmin46\arcsec$.  The
error radius is 0\farcm9 (at the 99\% confidence level). \bron\ is 0\farcm3
from this position and well within the WFC error circle. The nearest
other known X-ray source is 1RXS J171935.6-410054, at an angular
distance of 36\arcmin. We conclude that the flare is from \bron.

The WFC data do not show a detection of the persistent emission of
\bron\ over any part of the 1996 observations. The upper limit is
$6\times10^{-11}$ \ecs\ (2-10 keV), over all the 1996 data. The flux 
measured with ROSAT and translated to 2-10 keV numbers
(assuming a power-law spectrum with a photon index of 2.1 and $N_{\rm
H}=9.2\times10^{21}\rm~cm^{-2}$, see below) varies from
$1\times10^{-11}$ \ecs\ (September 1990) to $9\times10^{-12}$ \ecs\
(March 1994) to $4\times10^{-12}$ \ecs\ (September 1994).  The
uncertainty of these numbers is large due to the lack of knowledge of
the persistent spectrum, and we estimate it to be a factor of 2.

Unfortunately, the flare occurred close to earth occultation. Model
calculations predict that the flux in the heaviest degraded 
calibrated WFC-channel experiences absorption by less than 10\% as
long as the direction is higher than 145 km above the earth
limb. \bron\ sets behind the earth 209~s after the flare onset, while
the height of 145 km was reached after 135~s.

\section{Data analysis}
\label{analysis}
  
\begin{figure}[t]
\psfig{figure=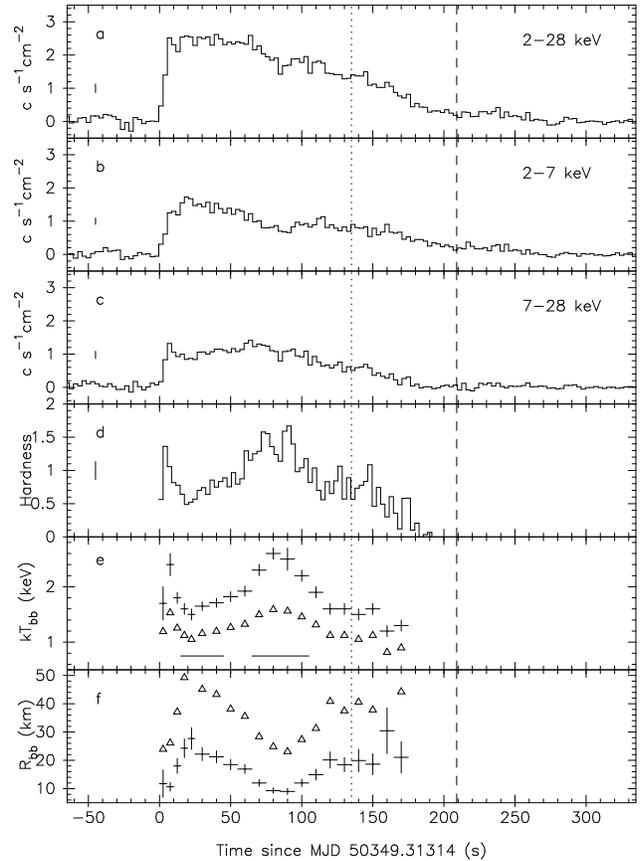,width=\columnwidth,clip=t}
\caption{{\bf a}, {\bf b} and {\bf c} show the light curve in the 2-28
keV, 2-7 keV and 7-28 keV energy bands. {\bf d} shows the hardness
ratio during the flare. The hardness ratio is the ratio of the count
rates in the 7-28 keV and the 2-7 keV bands. The bin time for {\bf
a-d} is 3 s, and typical error bars of length 1$\sigma$ for
respectively the fluxes and the hardness are indicated at the left.
The crosses in {\bf e} and {\bf f} show the black-body temperature
k$T_{\rm bb}$ (in keV) and the effective black-body radius $R_{\rm
bb}$ (at a distance of 10 kpc) of the radiating object during the
flare, respectively.  The triangles in {\bf e} and {\bf f} represent
the temperature and the radius corrected for scattering effects (see
Sect.~\ref{sec-nonplanck}).  The horizontal lines in {\bf e} indicate
the time intervals for which spectra are shown in
Fig.~\ref{fig:spectra}.  The dashed line marks the time when the
source sets behind the earth and the dotted line the time when the
source is below 145 km above the earth limb (see Sect.~\ref{detection}).
\label{fig:lcvs_kt_r_hard} }
\end{figure}

The light curve of the flare is shown in various passbands in the top
panels of Fig.~\ref{fig:lcvs_kt_r_hard}. The rise time to the peak
level is quick ($\sim$6 s), while the decay is comparatively slow
($\ga$ 200 s). During the first minute the flux is approximately
constant at a peak level of 2.6~ct~s$^{-1}$cm$^{-2}$ or 1.3 Crab units
(2-28 keV). The flare starts off at a relatively high hardness ratio
(see Fig.~\ref{fig:lcvs_kt_r_hard}d), after which it decreases during
$\sim$20~s, followed by an increase over $\sim$70~s to a similar peak
value. Finally, it decreases again for the remainder of the flare.

\begin{figure}[t]
\psfig{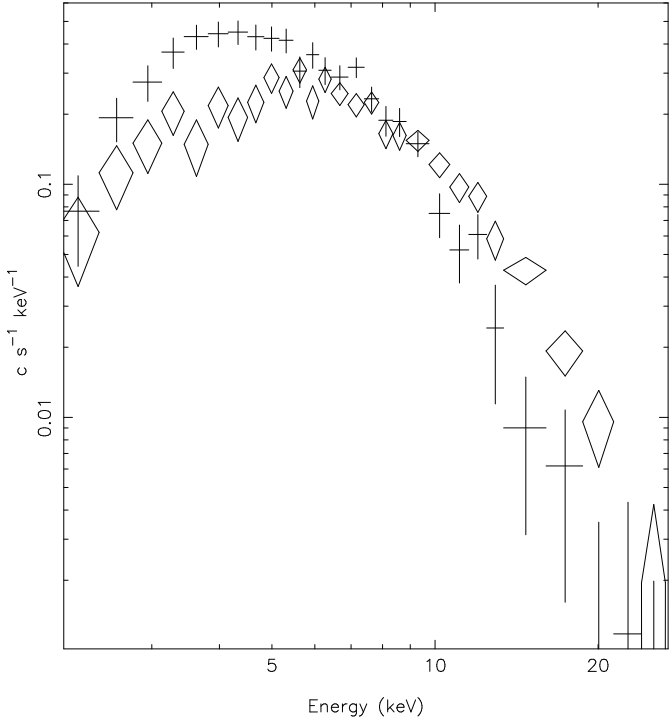}
\caption{Two examples of burst spectra from the time intervals indicated in 
Fig.~\ref{fig:lcvs_kt_r_hard}e. The crosses indicate the first interval and the
diamonds the second. The fitted blackbody temperatures are $1.63\pm0.04$ keV (crosses) 
and $2.38\pm0.07$ keV (diamonds).
\label{fig:spectra} }
\end{figure}

We modeled the flare spectrum with black-body radiation, thermal
bremsstrahlung and a power-law function, all subjected to interstellar
absorption. We used the data of the first and brightest 60 s of the
flare.  This duration is a compromise between obtaining sufficient
statistics for a meaningful model comparison, and limiting the
spectral change within the data set as suggested by the hardness
ratio. The best fits of the three models resulted in $\chi^{2}_{\rm
r}$ values of 0.72, 2.52 and 4.59 respectively, all for 26 degrees of
freedom. We conclude that black-body radiation is the best description
of the data.  During later parts of the flare, black-body fits also
gave the best results. Two examples of burst spectra are shown in
Fig.~\ref{fig:spectra}.

We carried out time-resolved spectroscopy by resolving the flare in 20
time intervals, generating spectra for those intervals, and fitting a
black-body model with absorption to each of those spectra. We kept the
absorbing column $N_{\rm H}$ tied to a single value over all intervals
up to the point where the source direction enters the earth atmosphere
(indicated by the dotted line in Fig.~\ref{fig:lcvs_kt_r_hard}) and
left it tied again in the group of spectra after that. The results of
the fit for the black body temperature k$T_{\rm bb}$ and radius
$R_{\rm bb}$ at an assumed distance of 10~kpc are presented in
Figs.~\ref{fig:lcvs_kt_r_hard}e and f. The errors in the fit
parameters are single-parameter 1$\sigma$ uncertainties. The absorbing
column over all intervals up to earth-atmosphere entry was found to be
$(3.1\pm1.3)\times10^{22}\rm~cm^{-2}$ (90\% confidence). This compares
to $9.2\times10^{21}\rm~cm^{-2}$ as found from interpolation of HI
maps (Dickey \& Lockman 1990), so is 3 times as high at about the
3$\sigma$ confidence level. This apparent inconsistency is, however,
not unexpected (see Arabadjis \& Bregman 1999).  During
earth-atmosphere entry the absorbing column was found to be
$(6\pm4)\times10^{22}\rm~cm^{-2}$.

After the peak of the flare the temperature of the emitting object is
decreasing. The radius plotted in Fig.~\ref{fig:lcvs_kt_r_hard}f is
that of a sphere with an area equal to that predicted by the
black-body flux, assuming that the source is at a distance of 10 kpc
and that the emission is isotropic. The radius peaks at the beginning
of the burst, after which it starts to decrease, reaching a minimum
after about 1 minute. It then increases again and levels off. There is
a strong anti-correlation between the temperature and the black body
radius.

\section{Discussion}
\label{discussion}
\subsection{Physical Nature}

Type-I X-ray bursts are believed to originate from thermonuclear
flashes on the surfaces of neutron stars and have three main
characteristics (see Lewin et al. 1993 for a review): 1) the shape is
commonly described by a fast rise ($<$ 10 s) and a longer exponential
decay, 2) the spectrum is best described by a black-body model and 3)
the spectrum shows softening during the decay phase (i.e. the neutron
star photosphere cools).  Some of the type-I bursts show evidence for
photospheric radius expansion. The luminosity in such a burst becomes
so high (i.e., reaches the Eddington limit) that the atmosphere of the
neutron star expands due to radiation pressure. When the luminosity
decreases below the Eddington limit, the atmosphere contracts again
until it reaches the neutron star surface.  During the expansion and
contraction phases, the luminosity will stay close to the
Eddington luminosity, because the excess luminosity is effectively
transformed into kinetic and potential energy of the atmosphere (Lewin
et al. 1993, and references therein). The profiles of these bursts do
not satisfy the first characteristic completely (fast rise and
exponential decay): an Eddington-limited burst can show a period of
almost constant flux after the rise, or a double peaked light
curve. In extreme cases a brief precursor can be seen to the main
event.

The flare from \bron\ shows characteristics of an Eddington-limited
type-I burst, see Fig.~\ref{fig:lcvs_kt_r_hard}. The hardness (and
equivalently the temperature) shows a peak in the beginning,
corresponding to the start of the burst and a following decrease due
to the expansion. When the atmosphere starts contracting, the
temperature rises again, until the radius reaches its original value,
after which the neutron star shows cooling, similar to type-I
bursts. The initial rise and the following contraction can be seen in
Fig.~\ref{fig:lcvs_kt_r_hard}e. Surprisingly, the radius starts to
increase again after it reaches a minimum value. This 
behavior is not unique and has been seen in other bursts (e.g. Tawara
et al. 1984, Hoffman et al. 1980, Van Paradijs et al. 1990). The
durations of those bursts were also long ($>$ 60 s). A possible
explanation is the deviation of the spectra from that of true
black-body radiation. This will be discussed in
Sect.~\ref{sec-nonplanck}.

If the burst is indeed Eddington-limited, it is possible to derive the
distance. The Eddington luminosity $L_{\rm Edd}$ for 1.4 $M_{\sun}$
neutron stars is about 2.0 $\times10^{38}$ erg s$^{-1}$. Note that
there is an uncertainty in this number because of the
uncertainties in the composition (and therefore the opacity) and the
mass of the neutron star. By making $L_{\rm Edd}$ equivalent to the
observed flux we find a distance of 6.5 $\pm$ 0.5 kpc, assuming that
the burst emission is isotropic, that the black-body temperature
corresponds to the effective temperature and taking into account
gravitational redshift effects.

Instead of the theoretical value for $L_{\rm Edd}$ we can use the
observed peak luminosity of Eddington-limited bursts seen in globular
clusters, for which the distances are known. This peak luminosity is
3.0 $\times10^{38}$ erg s$^{-1}$ (Lewin et al. 1993). From this we
find a distance of about 8 kpc. It is clear that our distance
determination to \bron\ is a rough estimate.

Assuming a distance of 6.5 kpc, the total energy emitted in the burst
is $\ga$2.3$\times10^{40}$ erg.  If the energy released per gram in
the nuclear fusion is between 1.6$\times10^{18}$ erg g$^{-1}$ and
6$\times10^{18}$ erg g$^{-1}$ (Lewin et al. 1995), a mass of at least
$3.8\times10^{21}$ g accreted material is involved. The persistent
emission is quite uncertain, but if the level is lower than 
the maximum ROSAT value, 1$\times 10^{-11}$ \ecs, the time
needed to accumulate $3.8\times10^{21}$ g is at least 150 days. From
the WFC upper limit we infer that the waiting time is at least 26
days. It is no surprise that no other bursts have been seen from 
this source at all.

The relatively low persistent emission can also explain the long
duration of the burst. At very low accretion rates, the flashing layer
is expected to be hydrogen rich (Fujimoto et al. 1981), and because of
the long time scales involved in proton-capture processes, bursts are
expected to last longer.  Observations confirm this anti-correlation
between low accretion rate and burst duration from a variety of burst
sources (Van Paradijs et al. 1988).

All known type-I X-ray bursters that have identified optical
counterparts are low-mass X-ray binaries (LMXBs, e.g., Van Paradijs
1995). In analogy, it is very likely that \bron\ is a LMXB as
well. Given the high level of Galactic absorption (predicting a severe
optical extinction of $A_{\rm V}\sim18$) and the relatively large
distance, it is difficult to find an optical counterpart.

\subsection{Non-Planckian Spectra} 
\label{sec-nonplanck}

In reality, X-ray bursts are not true black-body emitters. The spectra
resemble Planckian functions, but the colour temperatures derived from
these spectra are generally substantially higher than the true
effective temperatures (Van Paradijs et al. 1990, and references
therein). The reason is that electron scattering dominates the opacity
at high temperatures, causing the photons to be Comptonized to higher
energies. This has been confirmed by detailed modelling of neutron
star atmospheres (e.g. London et al.  1984, 1986, Ebisuzaki 1987,
Foster et al. 1986). These calculations show that the ratio of colour
temperature to effective temperature depends mainly on the ratio of
the luminosity to the Eddington luminosity. The dependency on other
parameters such as surface gravity and chemical composition is very
weak (see Lewin et al. 1993, and references therein).

To evaluate the magnitude of this effect in our burst, we made a
correction using analytical fits from Ebisuzaki and Nakamura (1988) to
the numerical results from Ebisuzaki (1987). According to these fits
the colour temperature can be twice as high as the effective
temperature if the luminosity is close to the Eddington
luminosity. The corrected temperature and radius are also shown in
Fig.~\ref{fig:lcvs_kt_r_hard}.  We assumed a hydrogen-rich composition
($X=0.73$).  It can be seen that the radius now increases a factor of
$\sim$1.6 after the minimum, i.e. less than the original increase (a
factor of $\sim$2).

Observations indicate that close to the Eddington limit the spectral
hardening may be even larger than suggested above (see Lewin et
al. 1993). If this is true, the increase of the radius will be even
less than the factor 1.6 derived above.

We verified the above theoretical approach by an empirical approach
obtained by Sztajno et al. (1985) from analysis of the source 4U
1636-53 and find similar results for k$T\la$2~keV.

\section{Conclusion}
\label{conclusion}
We have discovered an Eddington-limited thermonuclear burst from the
ROSAT source \bron. In analogy with other X-ray bursters, this
identifies this already-suspected X-ray binary as a low-mass X-ray
binary with a neutron star.  From equivalence of the peak flux and the
Eddington luminosity, we find a distance of 6.5~kpc.  ROSAT over the
years detected variable low-level persistent 0.1-2.4 keV X-ray
emission from this source which at maximum, given the distance,
translates to a 0.1-2.4 keV luminosity of 10$^{35}$~erg~s$^{-1}$. No
persistent emission was detected just before and after the burst above
an upper limit which is one order of magnitude larger than the ROSAT
measurements made several years prior to the burst.

\acknowledgement We thank Jaap Schuurmans and Gerrit Wiersma for WFC
software and data archival support. Paolo Soffitta is acknowledged for
his work on the earth atmosphere absorption model. The data reported
here were obtained during the Science Performance Verification phase.
The BeppoSAX satellite is a joint Italian and Dutch program.  This
research has made use of the Simbad database, operated at CDS,
Strasbourg, France, of data obtained through the High Energy
Astrophysics Science Archive Research Center Online Service, provided
by the NASA/Goddard Space Flight Center, and of the ROSAT Data Archive
of the Max-Planck-Institut f\"{u}r extraterrestrische Physik (MPE) at
Garching, Germany.

\end{document}